\documentclass[prd,twocolumn,showpacs,superscriptaddress,nofootinbib,floatfix,showkeys,10pt]{revtex4-2}
\usepackage{bm}
\usepackage{times}
\usepackage{braket}
\usepackage{amsfonts,amssymb,stmaryrd,latexsym,amsmath}
\usepackage[usenames,dvipsnames]{color}
\usepackage{epsfig}
\usepackage{slashed}
\usepackage{hyperref}
\usepackage{subfigure}
\usepackage{graphicx}
\usepackage{slashed}
\usepackage{orcidlink}
\usepackage{multirow}
\usepackage{appendix}
\usepackage{dashrule}

\allowdisplaybreaks[1]

\definecolor{nicered}{rgb}{0.7,0.1,0.1}
\definecolor{nicegreen}{rgb}{0.1,0.5,0.1}
\definecolor{emph}{rgb}{1,0,0}
\definecolor{doub}{rgb}{0.7,0.2,1.0}
\definecolor{navyblue}{RGB}{0, 110, 184}

\hypersetup{colorlinks,citecolor=nicegreen,linkcolor=nicered,urlcolor=navyblue}

 \newcommand{\clabel}[2][]{#2}
 
\begin{document}
%
	
	\title{Fully charmed P-wave tetraquark resonant states in the quark model} 
	\author{Wei-Lin Wu\,\orcidlink{0009-0009-3480-8810}}\email{wlwu@pku.edu.cn}
	\affiliation{School of Physics, Peking University, Beijing 100871, China}
	\author{Shi-Lin Zhu\,\orcidlink{0000-0002-4055-6906}}\email{zhusl@pku.edu.cn}
	\affiliation{School of Physics and Center of High Energy Physics,
		Peking University, Beijing 100871, China}
	
	\begin{abstract}
	We conduct the first comprehensive P-wave four-body dynamical calculations of the fully charmed tetraquark systems within the quark potential model. We apply the Gaussian expansion method to solve the four-body Schr\"odinger equation, incorporating both dimeson and diquark-antidiquark spatial configurations. The matrix elements of P-wave states are calculated analytically using the infinitesimally-shifted Gaussian basis functions.  With the complex scaling method, we obtain several fully charmed P-wave resonant states with compact tetraquark configuration in the mass region of $(7.0,7.2)$ GeV, including states with exotic quantum numbers $J^{PC}=0^{--},1^{-+}$. However, we find no resonant states with the mass $M<7$ GeV and width $\Gamma<200$ MeV. \clabel[Q3.4]{Combining the present investigation  with our previous results on S-wave fully charmed tetraquark systems~\cite{Wu:2024euj}, we find no candidates for the experimental states $X(6400)$ and $X(6600)$ in the quark model}.

	\end{abstract}
	
	\maketitle
	
	\section{Introduction}~\label{sec:intro}
	As novel hadronic states beyond conventional mesons $(q\bar q)$ and baryons $(qqq)$,  multiquark states exhibit more intricate internal structures, offering an excellent platform for studying the non-perturbative quantum chromodynamics (QCD). In the past few decades, hadron physicists have made significant progress on both the experimental and theoretical explorations of multiquark states~\cite{Chen:2016qju,Esposito:2016noz,Hosaka:2016pey,Lebed:2016hpi,Guo:2017jvc,Ali:2017jda,Brambilla:2019esw,Liu:2019zoy,Meng:2022ozq,Mai:2022eur,Chen:2022asf}.

    The fully charmed tetraquark states serve as an ideal and relatively pure multiquark system. Without the light degrees of freedom, it is unaffected by the complicated coupled-channel effects between the tetraquark states and $c\bar c$ mesons, as in the case of $X(3872)$. Moreover, the long-range light meson exchange mechanism is absent, making compact tetraquark states arising from short-range gluon exchange interactions more likely to exist. Several experimental candidates of the fully charmed tetraquark states have been reported in the $J/\psi J/\psi$ and $ J/\psi \psi(2S)$ decay channels. Specifically, $X(6900)$ was first discovered by the LHCb Collaboration~\cite{LHCb:2020bwg} and later confirmed by the CMS~\cite{CMS:2023owd} and ATLAS~\cite{ATLAS:2023bft}. $X(6400)$ was reported by the ATLAS~\cite{ATLAS:2023bft}, while $X(6600)$ and $X(7200)$ were reported by both the CMS~\cite{CMS:2023owd} and ATLAS~\cite{ATLAS:2023bft}. Numerous theoretical studies have discussed the experimentally observed fully charmed tetraquark states, offering various interpretations within different frameworks~\cite{Deng:2020iqw,liu:2020eha,Jin:2020jfc,Lu:2020cns,Chen:2020xwe,Albuquerque:2020hio,Giron:2020wpx,Wang:2020wrp,Dong:2020nwy,Zhang:2020hoh,Zhao:2020nwy,Gordillo:2020sgc,Zhang:2020xtb,Guo:2020pvt,Gong:2020bmg,Wan:2020fsk,Dosch:2020hqm,Yang:2020wkh,Zhao:2020jvl,   Faustov:2021hjs,Ke:2021iyh,Liang:2021fzr,Yang:2021hrb,Mutuk:2021hmi,Li:2021ygk,Wang:2021kfv,Dong:2021lkh,Wang:2021mma,Liu:2021rtn,Zhuang:2021pci,Asadi:2021ids,    Gong:2022hgd,Wang:2022jmb,Zhou:2022xpd,Wang:2022yes,Niu:2022cug,Dong:2022sef,Yu:2022lak,
    Ortega:2023pmr,Wang:2023jqs,Sang:2023ncm, Anwar:2023fbp,
    Wu:2024tif}. In our previous work~\cite{Wu:2024euj}, we carried out benchmark calculations of the S-wave fully charmed tetraquark systems within the quark potential model. We obtained compact tetraquark candidates for $ X(6900) $ and $ X(7200) $ in the $ J^{PC}=0^{++} $ and $ 2^{++} $ systems, but found no signals for $ X(6400) $ and $ X(6600) $. 
    
    It is natural to wonder whether parity negative P-wave fully charmed tetraquark states exist or not. Compared to the S-wave systems, P-wave tetraquark systems may contain more complex correlations among color, spin, and spatial degrees of freedom, due to the presence of various excitation modes and the inclusion of spin-orbit and tensor potentials. Investigations of the P-wave systems may enhance our understanding of exotic hadrons and enrich the hadron spectrum. The P-wave fully charmed tetraquark system was studied in the the QCD sum rule~\cite{Chen:2016jxd,Chen:2020xwe,Wang:2021mma,Chen:2024bpz}, coupled-channels formalism~\cite{Ortega:2023pmr}, diquark-antidiquark model~\cite{Debastiani:2017msn,Bedolla:2019zwg,Giron:2020wpx}, and the quark model with  diquark-antidiquark configuration~\cite{Deng:2020iqw,liu:2020eha,Wang:2021kfv,Yu:2022lak}.  However, the diquark-antidiquark picture does not consider the channel coupling effects with meson-meson continuum states and may lead to the misidentification of resonant states. 
	
	In this work, we conduct the first comprehensive P-wave four-body dynamical calculations of the fully charmed tetraquark systems within the quark potential model. We employ the Gaussian expansion method~\cite{HIYAMA2003223} and incorporate both dimeson and diquark-antidiquark spatial configurations to solve the four-body Schr\"odinger equation. The matrix elements of P-wave states are calculated analytically using the infinitesimally-shifted Gaussian basis functions. We utilize the complex scaling method~\cite{Aguilar:1971ve,Balslev:1971vb,aoyama2006complex} to distinguish genuine resonant states from meson-meson continuum states. We analyze the spatial structures of the resonant states by calculating their root-mean-square radii, which can clearly reveal whether they are meson molecules or compact tetraquark states. This framework has been successfully used to investigate S-wave tetraquark bound and resonant states~\cite{Meng:2023jqk,Chen:2023syh,Wu:2024euj,Wu:2024hrv,Wu:2024zbx}.
	
	This paper is organized as follows. In Sec.~\ref{sec:theo_framwork}, we introduce the theoretical framework, including the quark potential model, the wave function construction, the complex scaling method and the analysis of spatial structures. In Sec.~\ref{sec:result}, we present the numerical results and discussions. We summarize our findings in Sec.~\ref{sec:summary}.

	\section{Theoretical Framework}\label{sec:theo_framwork}
	
	\subsection{Hamiltonian}\label{subsec:hamiltonian}
	We employ the nonrelativistic quark potential model proposed in Ref.~\cite{Barnes:2005pb} to study the fully charmed tetraquark systems. The Hamiltonian in the center-of-mass frame reads
	\begin{equation}
		H=\sum_{i=1}^4 (m_i+\frac{p_i^2}{2 m_i})+\sum_{i<j=1}^4 V_{ij},
	\end{equation}
	with $m_i$ and $p_i$ being the mass and momentum of the (anti)quark $i$. The two-body potential $V_{ij}$, which results from the one-gluon-exchange and linear confinement interactions, consists of the central potential $V_{\rm cen}$, spin-orbit potential $V_{\rm so}$, and tensor potential $V_{\rm ts}$,
	\begin{equation}
		\label{eq:potential}
		\begin{aligned}
			&V_{i j} = -\frac{3}{16}\boldsymbol\lambda_i \cdot \boldsymbol\lambda_j(V_{\rm cen}(r_{ij}) + V_{\rm so}(r_{ij})+V_{\rm ts}(r_{ij})),\\
			&V_{\rm cen}(r_{ij}) = -\frac{4}{3}\frac{\alpha_s}{r_{ij}}+br_{ij}+\frac{32\pi\alpha_s}{9m_c^2}\delta^{(3)}(\boldsymbol{r}_{ij})\boldsymbol{S}_i\cdot\boldsymbol{S}_j,\\
			&V_{\rm so}(r_{ij})=\frac{1}{m_c^2}\left(\frac{2\alpha_s}{r_{ij}^3}-\frac{b}{2r_{ij}}\right)\boldsymbol{L}_{ij}\cdot(\boldsymbol{S}_i+\boldsymbol{S}_j),\\
			&V_{\rm ts}(r_{ij})= \frac{4\alpha_s}{3m_c^2}\frac{1}{r_{ij}^3}(\frac{3\left(\boldsymbol{S}_i \cdot \boldsymbol{r}_{i j}\right)\left(\boldsymbol{S}_j \cdot \boldsymbol{r}_{i j}\right)}{r_{ij}^2}-\boldsymbol{S}_i \cdot \boldsymbol{S}_j ),
		\end{aligned}
	\end{equation}
	where $\boldsymbol\lambda_i$ and $ \boldsymbol{S}_i $ are the $\mathrm{SU}(3)$ color Gell-Mann matrix and the spin operator acting on (anti)quark $ i $, respectively. $ r_{ij} $ is the distance between (anti)quark $ i $ and $ j $, and $\boldsymbol{L}_{ij}$ is the relative orbital angular momentum operator. 
	
	In the Schr\"odinger equation, the $\delta^{(3)}(\boldsymbol{r})$ and $\frac{1}{r^3}$ terms have singularities, which are artifacts of the nonrelativistic approximation. To avoid the singularities, the former is regularized by a Gaussian smearing function,
	\begin{equation}
		\delta^{(3)}(\boldsymbol{r})\rightarrow\frac{\sigma^3\exp{(-\sigma ^2r^2)}}{\pi^{3/2}},
	\end{equation} 
	while the $\frac{1}{r^3}$ terms in $V_{\rm so}$ and $V_{\rm ts}$ are often treated as perturbations in literature~\cite{Barnes:2005pb,Wang:2021kfv}. In this work we treat both singular terms in a consistent way by regularizing the $\frac{1}{r^3}$ terms as follows,
	\begin{equation}\label{eq:reg1/r3}
		\frac{1}{r^3}\rightarrow \frac{(1-e^{-\sigma_1^2r^2})^2}{r^3}.
	\end{equation}
	Instead of treating $V_{\rm so}$ and $V_{\rm ts}$ as perturbations, we solve the Schr\"odinger equation by diagonalizing the complete Hamiltonian with the regularized potential.  The parameters for the potential are listed in Table~\ref{tab:para}. The first four parameters, which are taken from Ref.~\cite{Barnes:2005pb}, were determined by fitting the charmonium spectrum. The last parameter $\sigma_1$ is determined by fitting the $\chi_{cJ}$ spectrum. Both the parameters $\sigma$ and $\sigma_1$ are comparable to $m_c$, which are reasonable for regularization to the nonrelativistic approximation. The theoretical masses of the charmonia as well as their root-mean-square (rms) radii are listed in Table~\ref{tab:meson}. Both the results from the regularized potential and the perturbation method are consistent with the experimental values within tens of MeV. Accordingly, we expect the uncertainties of the tetraquark spectrum to be of the same order.
	\begin{table}[htbp]
		\centering
		\caption{The parameters in the potential.}
		\label{tab:para}
		\begin{tabular*}{\hsize}{@{}@{\extracolsep{\fill}}ccccc@{}}
			\hline\hline
			$\alpha_s$ &$b{ [\mathrm{GeV}^{2}]}$&$m_c{ [\mathrm{GeV}]}$&$\sigma{ [\mathrm{GeV}]}$&$\sigma_1{ [\mathrm{GeV}]}$\\
			\hline
			0.5461&0.1425&1.4794&1.0946&1.3133\\
			\hline\hline
		\end{tabular*}
	\end{table}
	
	\begin{table}
		\centering
		\caption{The theoretical masses $m_{\rm Theo.}$ (in  $\mathrm{MeV}$) of charmonia using the regularized potential, compared with the experimental results $m_{\rm Exp.}$ taken from Ref.~\cite{ParticleDataGroup:2024cfk}, and the theoretical results $m_{\rm Per.}$ in Ref.~\cite{Barnes:2005pb} using the perturbation method. The rms radii (in fm) are listed in the last column. }
		\label{tab:meson}
		\begin{tabular*}{\hsize}{@{}@{\extracolsep{\fill}}ccccc@{}}
			\hline\hline
			Mesons& $ m_{\rm Exp.} $&$ m_{\rm Theo.} $& $m_{\rm Per.}$&$ r^{\rm rms}_{\rm Theo.} $ \\
			\hline
			$\eta_c$ & 2984 &2982&2982&0.36\\
			$\eta_c(2S)$ & 3638 &3630&3630&0.83 \\ 
			$J/\psi$ & 3097 &3090&3090&0.41 \\
			$\psi(2S)$ & 3686 &3672&3672&0.86 \\ 
			$\psi(3S)$ &4040 &4072&4072&1.24 \\
			$ \chi_{c0} $&3415&3426&3424&0.59\\
			$ \chi_{c1} $&3511&3511&3505&0.67\\
			$ \chi_{c2} $&3556&3544&3556&0.71\\
			$ h_c $&3525&3516&3516&0.67\\
			$ \chi_{c0}(2P) $&$\cdots$&3876&3852&1.01\\
			$ \chi_{c1}(2P) $&$\cdots$&3933&3925&1.07\\
			$ \chi_{c2}(2P) $&$\cdots$&3959&3972&1.10\\
			$h_c(2P)$&$\cdots$&3934&3934&1.07\\
			\hline\hline
			
			\hline\hline
		\end{tabular*}
	\end{table}
	 
	\subsection{Wave function }\label{subsec:wf}
	The P-wave tetraquark wave functions are expanded using a set of basis functions, which are written as
	\begin{equation}
		\psi^J=\mathcal{A}(\chi^S\otimes\Phi^{L=1})^J,
	\end{equation}
	where $ \mathcal{A} $ is the antisymmetric operator of identical particles, $ \phi^{L=1} $ and $ \chi^S $ represent the P-wave spatial wave function and color-spin wave function with total spin $S$, respectively. The total angular momentum $J=L{\oplus} S$. A complete set of the color-spin wave function is used,
	\begin{equation}\label{eq:colorspin_wf}
		\begin{aligned}
			\chi^{S,s_1,s_2}_{\bar 3_c\otimes 3_c}=\left[\left(Q_1Q_2\right)_{\bar 3_c}^{s_1}\left(\bar q_1\bar q_2\right)_{3_c}^{s_2}\right]_{1_c}^{S},\\
			\chi^{S,s_1,s_2}_{6_c\otimes \bar 6_c}=\left[\left(Q_1Q_2\right)_{6_c}^{s_1}\left(\bar q_1\bar q_2\right)_{\bar{6}_c}^{s_2}\right]_{1_c}^{S},\\
		\end{aligned}
	\end{equation}
	for all possible combinations of $ S, s_1, s_2 $. The subscripts and superscripts denote the color and spin representations, respectively. 
	
\begin{figure}[htbp]
		\centering
		\includegraphics[width=.9\linewidth]{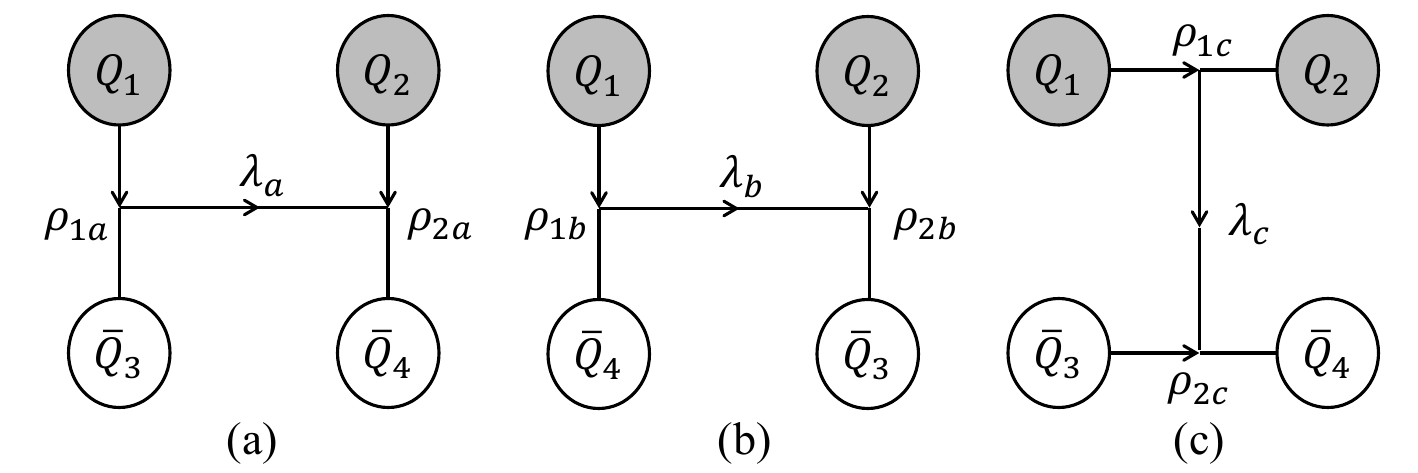}
		\caption{The Jacobian coordinates for two types of spatial configurations: (a), (b) for the dimeson configurations, and (c) for the diquark-antidiquark configuration.}
		\label{fig:jac}
	\end{figure}

	For the spatial wave function, we employ the Gaussian expansion method~\cite{HIYAMA2003223} and consider three sets of spatial configurations (dimeson and diquark-antidiquark), which are denoted by $ (\rm{jac}) = (a),\,(b),\,(c) $. In each configuration, there are three independent Jacobian coordinates $\rho_{1}, \rho_{2}, \lambda$, as shown in Fig.~\ref{fig:jac}. We consider one orbital excitation, i.e., the $ \rho $ mode excitation $ |l_{\rho_1}=1, l_{\rho_2}=l_\lambda=0\rangle $ or $ |l_{\rho_1}=0, l_{\rho_2}=1, l_\lambda=0\rangle $, and the $ \lambda $ mode excitation $ |l_{\rho_1}=l_{\rho_2}=0,l_\lambda=1\rangle $. The spatial basis functions are written as 

	\begin{equation}\label{eq:spatial_wf}
		\begin{aligned}
		\Phi^{L=1}_{n_{1},n_{2},n_{3}} = \,&\phi_{n_1l_{\rho_1}}(\rho_{1})\phi_{n_{2}l_{\rho_2}}(\rho_{2})\phi_{n_{3}l_\lambda}(\lambda)\cdot\\
		&\left[\left(Y_{l_{\rho_1}}(\hat{\rho_1})\otimes Y_{l_{\rho_2}}(\hat{\rho_2})\right)\otimes Y_{l_\lambda}(\hat{\lambda})\right]^{L=1}.
		\end{aligned}
	\end{equation}
	 $ \phi_{nl}(r) $ takes the Gaussian form,
	\begin{equation}
		\begin{array}{c}
			\phi_{nl}(r)=N_{nl}r^le^{-\nu_{n}r^2},\\
			\nu_{n}=\nu_{1}\gamma^{n-1}\quad (n=1\sim n_{\rm max}),
		\end{array}
	\end{equation}
	where $ N_{nl} $ is the normalization factor. To obtain numerically stable results, we take $n_{\rm max}=10\sim12$, and the total number of bases is $10^4\sim10^5$. 
	
	To evaluate the spatial matrix elements of P-wave states, we make use of the infinitesimally-shifted Gaussian basis functions~\cite{HIYAMA2003223}, which circumvent the need for carrying out complicated integrals of spherical harmonics of different configurations. We present the details in Appendix~\ref{appsec:me}.

	\subsection{Complex scaling method}
	The wave functions of resonant states are not square integrable, causing challenges for solving the Schr\"odinger equation numerically. To obtain possible bound states and resonant states, we employ the complex scaling method (CSM)~\cite{Aguilar:1971ve,Balslev:1971vb,aoyama2006complex}, where the coordinate $ \boldsymbol{r} $ and its conjugate momentum $ \boldsymbol{p} $ are transformed as
	\begin{equation}
    \label{eq:csmtrans}
		U(\theta) \boldsymbol{r}=\boldsymbol{r} e^{i \theta}, \quad U(\theta) \boldsymbol{p}=\boldsymbol{p} e^{-i \theta}.
	\end{equation}
	Under such a transformation, the Hamiltonian is analytically continued to the complex energy plane and no longer Hermitian,
	\begin{equation}
		H(\theta)=\sum_{i=1}^4 (m_i+\frac{p_i^2e^{-2i\theta}}{2 m_i})+\sum_{i<j=1}^4 V_{ij}(\boldsymbol{r}_{ij}e^{i\theta}).
	\end{equation}
	\clabel[Q1.1]{The wave function is transformed as~\cite{aoyama2006complex}
    \begin{equation}
    \label{eq:csmwftrans}
        \Psi^\theta=U(\theta)\Psi=e^{3i\theta/2}\Psi(re^{i\theta}),
    \end{equation} 
    and its asymptotic behavior is modified accordingly.} If the complex scaling angle $ \theta\in[0,\frac{\pi}{2}) $, the wave functions of bound states remain square integrable. They are located on the negative real axis in the energy plane.  If  $ \theta>\theta_r=\frac{1}{2}\tan^{-1}(\Gamma_r/2M_r) $, where $ M_r $ and $ \Gamma_r $ represent its mass and width, the wave functions of the resonant states become square integrable and are obtained with eigenenergy $ E_r=M_r-i\Gamma_r/2 $. Scattering states align along rays starting from threshold energies with $ \operatorname{Arg}(E)=-2\theta $. By choosing appropriate complex scaling angles and solving the complex scaled Schr\"odinger equation, we can obtain bound states, resonant states and scattering states simultaneously.
	\subsection{Spatial structures}\label{subsec:spatial_stru} 
	Generally, tetraquark states are classified into meson molecules and compact tetraquarks. \clabel[Q4]{The molecular and compact configurations are often treated in the hadronic level~\cite{Meng:2019ilv,Lin:2024qcq} and the diquark-antidiquark picture~\cite{Jaffe:2004ph,Maiani:2004vq}, respectively.}  Both configurations are allowed in the quark model, where no assumptions regarding the clustering of quarks are made. In our previous works~\cite{Chen:2023syh,Wu:2024euj}, we have proposed to use the rms radii to analyze the spatial structures of tetraquark states. We decompose the complete antisymmetric wave function as
	
	\begin{equation}\label{eq:wf_decompose}
		\begin{aligned}
			\Psi^\theta=&\sum_{S,s_1\geq s_2}\left([(Q_1\bar Q_3)^{s_1}_{1_c}(Q_2\bar Q_4)^{s_2}_{1_c}]^{S}_{1_c}\otimes|\psi_1^{S,s_1s_2}\rangle\right.\\
			&+[(Q_1\bar Q_3)^{s_2}_{1_c}(Q_2\bar Q_4)^{s_1}_{1_c}]^{S}_{1_c}\otimes|\psi_2^{S,s_1s_2}\rangle\\
			&+[(Q_1\bar Q_4)^{s_1}_{1_c}(Q_2\bar Q_3)^{s_2}_{1_c}]^{S}_{1_c}\otimes|\psi_3^{S,s_1s_2}\rangle\\
			&+\left.[(Q_1\bar Q_4)^{s_2}_{1_c}(Q_2\bar Q_3)^{s_1}_{1_c}]^{S}_{1_c}\otimes|\psi_4^{S,s_1s_2}\rangle\right)\\
			=&\mathcal{A}\sum_{S,s_1\geq s_2}[(Q_1\bar Q_3)^{s_1}_{1_c}(Q_2\bar Q_4)^{s_2}_{1_c}]^{S}_{1_c}\otimes|\psi_1^{S,s_1s_2}\rangle\\
			\equiv&\mathcal{A}\,\Psi_{\rm nA}^\theta,
		\end{aligned}
	\end{equation}
	Instead of using the complete wave function $ \Psi^\theta $, we use the decomposed non-antisymmetric wave function $ \Psi_{\mathrm{nA}}^\theta $ to calculate the rms radius:
	\begin{equation}\label{eq:rmsr}
		r^{\mathrm{rms}}_{ij}\equiv \mathrm{Re}\left[\sqrt{\frac{\langle\Psi_{\mathrm{nA}}^\theta | r_{ij}^2 e^{2i\theta}|\Psi_{\mathrm{nA}}^\theta\rangle}{\langle\Psi_{\mathrm{nA}}^\theta | \Psi_{\mathrm{nA}}^\theta\rangle}}\right].
	\end{equation}
	
	This definition of the rms radius has been successfully used to distinguish between meson molecules and compact tetraquarks in various systems~\cite{Chen:2023syh,Wu:2024euj,Wu:2024hrv,Ma:2024vsi,Wu:2024zbx}. For a meson molecule, $Q_1\bar Q_3$ and $Q_2\bar Q_4$ form two color singlets in $ \Psi_{\mathrm{nA}}(\theta) $, so $ r^{\mathrm{rms}}_{13} $ and $ r^{\mathrm{rms}}_{24} $ are expected to be the sizes of the constituent mesons, and much smaller than the other rms radii. For a compact tetraquark state, all four (anti)quarks are confined together, and all rms radii are on the order of $\Lambda_{QCD}^{-1}\sim$ 1 fm. It is worth noting that the rms radii obtained from the conventional definition using the complete wave function $ \Psi(\theta) $  and those from our novel definition are qualitatively consistent for compact tetraquarks, as discussed in Refs.~\cite{Wu:2024euj,Wu:2024zbx}.

    \clabel[Q1.2]{It should be stressed that the inner products in the CSM are defined using the c-product~\cite{ROMO1968617,Moiseyev:1998gjp},
    \begin{equation}
	   \langle\phi_n \mid \phi_m\rangle\equiv\int \phi_n(r)\phi_{m}(r)d^3r,
    \end{equation}
    where the ``bra" states are identical to the ``ket" states, rather than their complex conjugates. Using the c-product and the transformation behavior of the wave function in Eq.~\eqref{eq:csmwftrans}, it can be shown that the rms radius defined in Eq.~\eqref{eq:rmsr} is independent of the complex scaling angle $\theta$, as expected for a physical quantity. However, in practical numerical calculations, we only use a finite set of basis functions to solve the complex scaled Schr\"odinger equation. Some of the resonant states may be less numerically stable, and their rms radii change as $\theta$ varies. In these scenarios, we cannot determine their configurations in the current calculations.
    }
    
	\section{Results and Discussions}\label{sec:result}
    \begin{figure*}[tbp]
		\centering
		\includegraphics[width=0.85\linewidth]{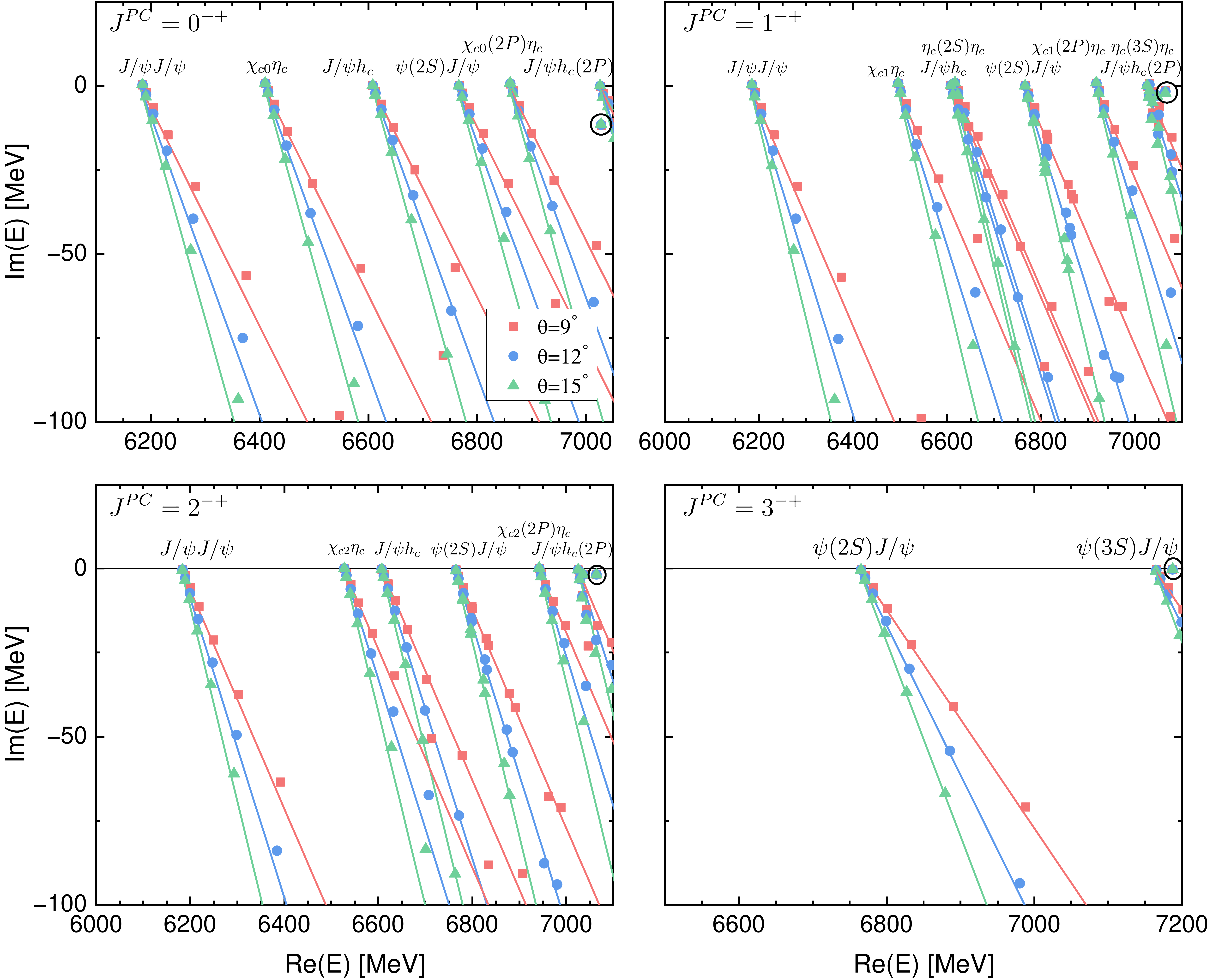}
		\caption{The complex energy eigenvalues of C-parity positive $cc\bar c\bar c$ states with varying $\theta$ in the CSM. The solid lines represent the continuum lines rotating along $\operatorname{Arg}(E)=-2 \theta$. The resonances do not shift as $\theta$ changes and are highlighted by the circles.}
		\label{fig:c+}
	\end{figure*}
	\begin{figure*}[tbp]
		\centering
		\includegraphics[width=0.85\linewidth]{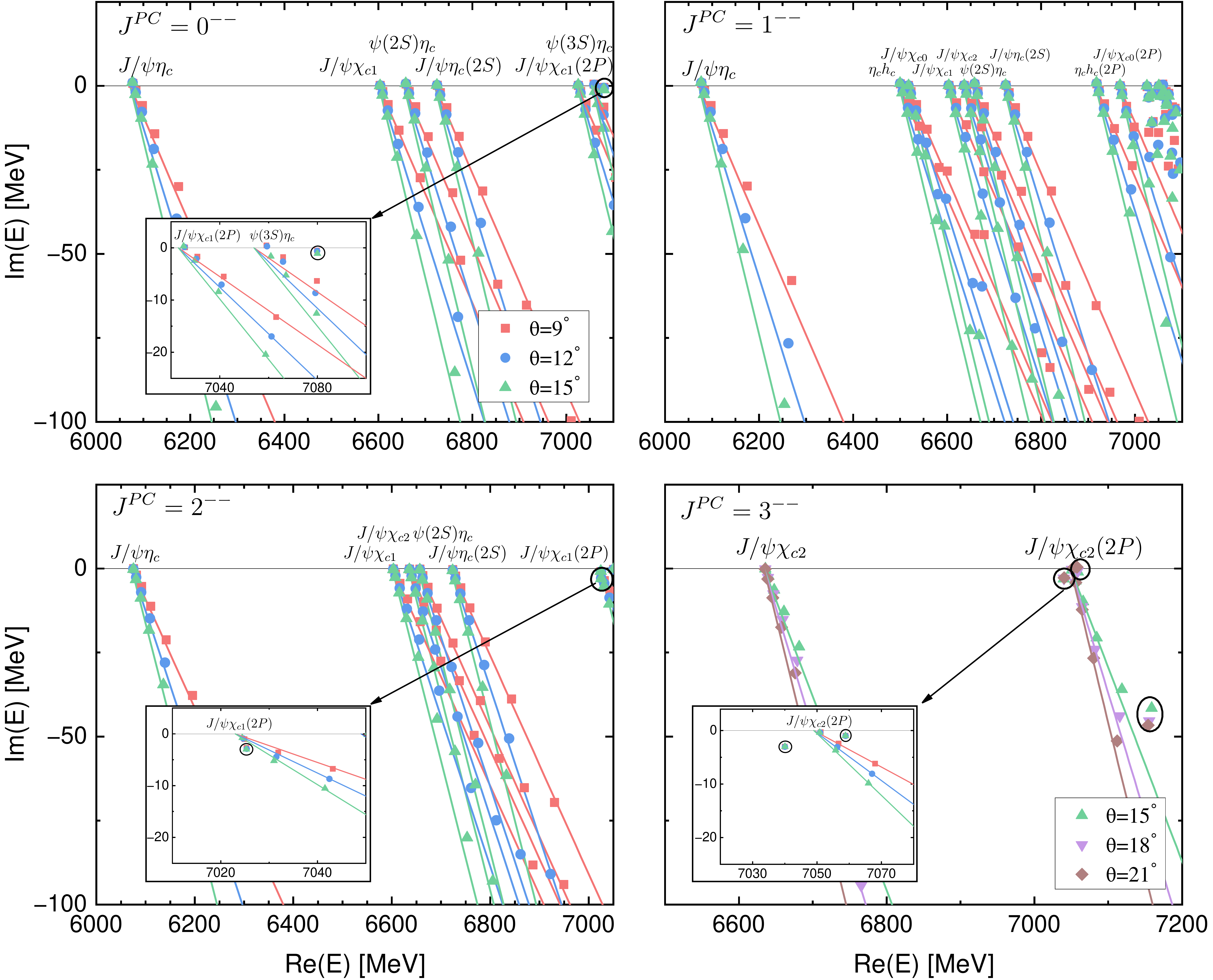}
		\caption{The complex energy eigenvalues of C-parity negative $cc\bar c\bar c$ states with varying $\theta$ in the CSM. The solid lines represent the continuum lines rotating along $\operatorname{Arg}(E)=-2 \theta$. The resonances do not shift as $\theta$ changes and are highlighted by the circles.}
		\label{fig:c-}
	\end{figure*}
	
	We investigate the P-wave fully charmed tetraquark systems with all possible quantum numbers, including $J^{PC}=0^{-+},0^{--},1^{-+},1^{--},2^{-+},2^{--},3^{-+},3^{--}$. We choose varying complex scaling angles $\theta$ in the CSM to calculate the complex energy spectra. The results for C-parity positive and negative systems are shown in Figs.~\ref{fig:c+} and \ref{fig:c-}, respectively. All of the states lie above the lowest dimeson threshold and no P-wave tetraquark bound state is found. We obtain meson-meson scattering states of all possible channels and several resonant states, which are labeled as $T_{4c,J^{PC}}(M)$ with $M$ being the mass of the state. The resonant states are found in the mass region of $ (7.0,7.2) $ GeV. They are located near the $ M(1S)M'(2P) $ thresholds, except in the $ 3^{-+} $ system where no such dimeson threshold exists, and $ T_{4c,3^{-+}}(7187) $ lies above the $ \psi(3S)J/\psi $ threshold. In the $1^{--}$ system, resonant states above $7$ GeV might exist, but we fail to identify them since the dimeson thresholds in the region are nearly degenerate, making it very difficult to identify resonant states from continuum states unambiguously in the current calculations.  
	
	The complex energies, proportions of different color and spin configurations and rms radii of the resonant states are summarized in Table~\ref{tab:structure}. The resonant states are mixtures of $\chi_{\bar{3}_c\otimes3_c}$ and $\chi_{6_c\otimes \bar6_c}$ configurations, suggesting that the color configuration mixing effect plays an important role. On the contrary, the mixing effect between different spin configurations is small, and the resonant state is dominated by one single $\chi^S$ component. This is expected since the spin mixing effect can only be induced by spin-orbit and tensor potentials, which are high-order corrections with small contributions.
	
	The rms radii of the resonant states are on the order of $ \Lambda_{QCD}^{-1}\sim 1 $ fm, indicating that they are all compact tetraquark states. Based on the classifications of compact tetraquarks proposed in Ref.~\cite{Wu:2024zbx}, $T_{4c,0^{-+}}(7028)$ is a compact diquark-antidiquark tetraquark, as its rms radii $r_{c_1c_2}^{\mathrm{rms}}, r_{\bar{c}_3\bar{c}_4}^{\mathrm{rms}}$ are relatively small. The other resonant states are compact even tetraquarks, where the relative distances between four (anti)quarks are of similar size. The rms radii for $T_{4c,2^{--}}(7025)$ changes drastically as $\theta$ varies, which may result from the fact that the state lies very close to the dimeson threshold and is strongly coupled to the scattering states. We cannot obtain numerically stable results and determine its configuration in the current calculations.

	\begin{table*}[tbp]
		\centering
		\caption{The complex energies (in MeV), proportions of different color and spin configurations and rms radii (in fm) of the $ cc\bar c\bar c $ resonant states. The ``?" indicates that the rms radii results are numerically unstable as the complex scaling angle $ \theta $ varies.}
		\label{tab:structure}
		\begin{tabular*}{\hsize}{@{}@{\extracolsep{\fill}}cc|cc|ccc|cccc@{}}
			\hline\hline
			$ J^{PC}$& $ M-i\Gamma/2 $ & \multicolumn{2}{c|}{$ \chi_{\rm color} $}&\multicolumn{3}{c|}{$ \chi^{\rm spin} $}&\multicolumn{4}{c}{$ r^{\rm rms} $}\\
			&&$ \chi_{\bar{3}_c\otimes3_c} $ &$ \chi_{6_c\otimes \bar6_c} $& $ \chi^{S=0} $&$ \chi^{S=1} $&$ \chi^{S=2} $&$ r_{c_1\bar{c}_3}^{\mathrm{rms}} $&$ r_{c_2\bar{c}_4}^{\mathrm{rms}} $&$ r_{c_1\bar{c}_4}^{\mathrm{rms}} = r_{c_2\bar{c}_3}^{\mathrm{rms}} $&$ r_{c_1c_2}^{\mathrm{rms}} = r_{\bar{c}_3\bar{c}_4}^{\mathrm{rms}} $\\
			\hline
			$0^{-+}$&${7028-12i}$&$ 68\% $&$ 32\% $& - &$ 100\% $&-&$ 0.75 $&$ 0.96 $&$ 0.69 $&$ 0.47 $\\
			$ 0^{--} $&$ 7080-1i $&$ 20\% $&$ 80\% $& - &$ 100\% $&-&$ 0.74 $&$ 0.73 $&$ 0.84 $&$ 0.95 $\\
			$1^{-+}$&$7065-2i$&$ 24\% $&$ 76\% $& $0\%$ &$ 100\% $&$ 0\% $&$ 0.66 $&$ 0.77 $&$ 0.88 $&$ 0.94 $\\
			$ 2^{-+} $&$ 7064-2i $&$ 27\% $&$ 73\% $& - &$ 100\% $&$ 0\% $&$ 0.67 $&$ 0.78 $&$ 0.84 $&$ 0.86 $\\
			$ 2^{--} $&$ {7025-3i} $&$ 73\% $&$ 27\% $& - &$ 5\% $&$ 95\% $&$ 0.87 $&$ 0.87 $&$ ? $&$ ? $\\
			$ 3^{-+} $&$ 7187-0.3i $&$ 85\% $&$ 15\% $& - & - &$ 100\% $&$ 0.82 $&$ 0.82 $&$ 0.84 $&$ 0.79 $\\
			$ 3^{--} $&$ 7040-3i $&$ 77\% $&$ 23\% $& - & - &$ 100\% $&$ 0.89 $&$ 0.89 $&$ 1.04 $&$ 0.85 $\\
			&$ 7059-1i $&$ 45\% $&$ 55\% $& - & - &$ 100\% $&$ 0.75 $&$ 0.75 $&$ 1.00 $&$ 0.86 $\\
			&$ 7154-46i $&$ 72\% $&$ 28\% $& - & - &$ 100\% $&$ 0.95 $&$ 0.95 $&$ 1.03$&$ 0.95 $\\
			
			\hline\hline
		\end{tabular*}
	\end{table*}

	The resonant states have various decay channels, as shown in Figs.~\ref{fig:c+} and \ref{fig:c-}. 
    Their S-wave and P-wave strong decay modes are summarized in Table~\ref{tab:decay}.  Among these states, $T_{4c,0^{--}}(7080)$ and $T_{4c,1^{-+}}(7065)$ are of particular interest. They have exotic quantum numbers which are not accessible by $c\bar c$ mesons. The former can be searched for in the S-wave $J/\psi\chi_{c1}$ and P-wave $J/\psi\eta_c$ channels, while the latter can be searched for in the S-wave $\chi_{c1}\eta_c,$  $J/\psi h_c $,  and P-wave $ J/\psi J/\psi $ channels. $ T_{4c,3^{-+}}(7187) $ has the smallest width $ 0.6 $ MeV, which may be due to the fact that it has no S-wave decay mode and can only decay into the $\psi(2S)J/\psi$ and $\psi(3S)J/\psi$ channels via P-wave transitions. It should be noted that the theoretical widths of the tetraquark resonant states are underestimated, since the widths of charmonia are not taken into account. Moreover, other decay modes via the annihilation of charmed quarks are also not considered in the quark model.
    
	 \begin{table*}[tbp]
        \centering
		\caption{The S-wave and P-wave {two-body} strong decay modes of the $ cc\bar c\bar c $ resonant states. }
		\label{tab:decay}
		\begin{tabular*}{\hsize}{@{}@{\extracolsep{\fill}}ccc@{}}
			\hline\hline
			States& S-wave & P-wave\\
			\hline
            $ T_{4c.0^{-+}}(7028) $& $ \chi_{c0}\eta_c,$ $ J/\psi h_c,$ $ \chi_{c0}(2P)\eta_c, $ $ J/\psi h_c(2P) $&$ J/\psi J/\psi,$ $ \psi(2S)J/\psi$\\
            $ T_{4c,0^{--}}(7080) $&$ J/\psi \chi_{c1},$ $ J/\psi\chi_{c1}(2P)$&$ J/\psi\eta_c,$ $ \psi(2S)\eta_c,$ $ J/\psi\eta_c(2S), $ $ \eta_c\psi(3S) $\\
            $T_{4c,1^{-+}}(7065)$&$\chi_{c1}\eta_c, $ $J/\psi h_c,$ $\chi_{c1}(2P)\eta_c,$ $J/\psi h_c(2P)$& $J/\psi J/\psi,$ $\eta_c(2S)\eta_c,$ $\psi(2S)J/\psi,$ $\eta_c(3S)\eta_c$\\
            $ T_{4c,2^{-+}}(7064) $&$ \chi_{c2}\eta_c,$ $ J/\psi h_c,$  $ \chi_{c2}(2P)\eta_c, $ $ J/\psi h_c(2P)$& $ J/\psi J/\psi,$ $ \psi(2S)J/\psi$ \\
            $ T_{4c,2^{--}}(7025) $&$ J/\psi \chi_{c1},$ $ J/\psi \chi_{c2},$ $ J/\psi\chi_{c1}(2P)$&$ J/\psi\eta_c,$ $ \psi(2S)\eta_c,$ $ J/\psi\eta_c(2S)$ \\
            $ T_{4c,3^{-+}}(7187) $&-& $ \psi(2S)J/\psi,$ $ \psi(3S)J/\psi $\\
            $ T_{4c,3^{--}}(7040) $&$ J/\psi\chi_{c2} $&-\\
            $ T_{4c,3^{--}}(7059) $&$ J/\psi\chi_{c2},J/\psi\chi_{c2}(2P) $&-\\
            $ T_{4c,3^{--}}(7154) $ &$ J/\psi\chi_{c2},J/\psi\chi_{c2}(2P) $&-\\
            
			\hline\hline
		\end{tabular*}
    \end{table*}

\begin{figure*}[tbp]
		\centering
		\includegraphics[width=\linewidth]{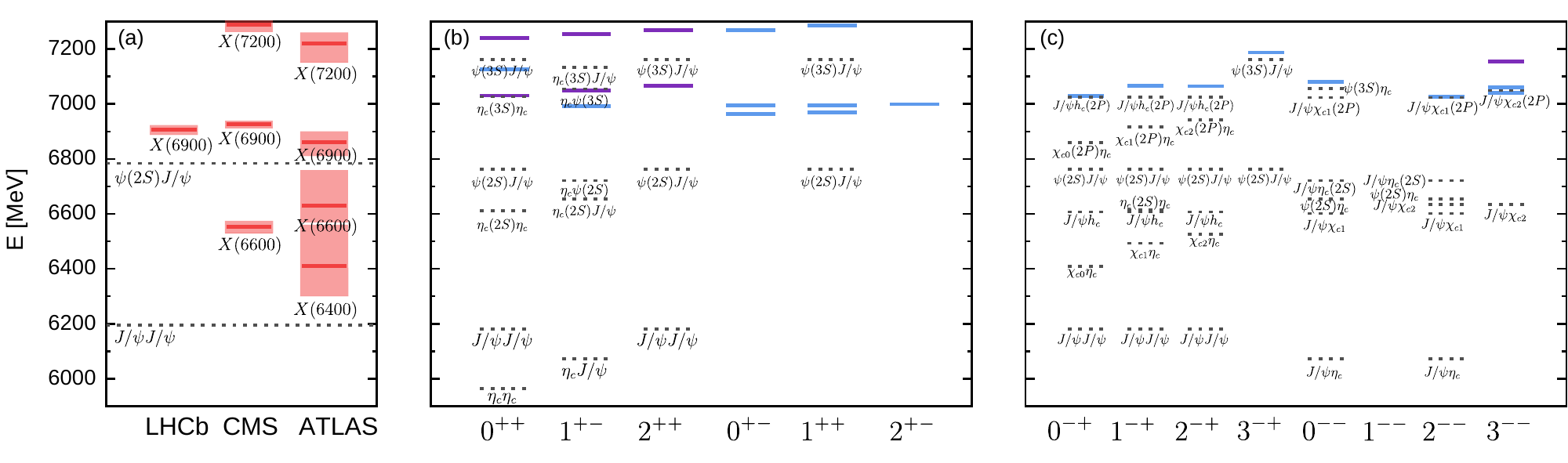}
		\caption{The experimental and theoretical mass spectrum of the $ cc\bar c\bar c $ states. The experimental results reported by LHCb (model \uppercase\expandafter{\romannumeral1})~\cite{LHCb:2020bwg}, CMS (noninterference model)~\cite{CMS:2023owd} and ATLAS (model A and $ \alpha $)~\cite{ATLAS:2023bft} are shown in Fig.~(a). The red lines and the dashed areas represent the central masses and the uncertainties in the experiments, respectively. The theoretical results of S-wave~\cite{Wu:2024euj} and P-wave $cc\bar c\bar c$ states in the quark potential model are shown in Fig.~(b) and (c). The purple and blue lines represent the theoretical states with widths larger and smaller than $ 30\,\mathrm{MeV} $, respectively. The blue states are rather narrow and do not match well with experimentally observed states. The dotted lines represent di-charmonium decay channels, whose experimental energies are taken from Ref.~\cite{ParticleDataGroup:2024cfk}.}
		\label{fig:spectrum4c}
	\end{figure*}

    \clabel[Q3.1]{In Fig.~\ref{fig:spectrum4c}, we compare the theoretical mass spectrum of the S-wave and P-wave fully charmed tetraquark states with the experimental results. The P-wave states are obtained in this work and the S-wave states were taken from our previous work~\cite{Wu:2024euj} using the same quark potential model~\cite{Barnes:2005pb}. In S-wave systems, good candidates for $ X(6900) $ and $ X(7200) $ are obtained in the $ J^{PC}=0^{++} $ and $ 2^{++} $ systems. Their masses and widths are comparable with the experimental results. However, the widths of the P-wave resonant states that can decay into the experimentally observed modes ($J/\psi J/\psi$ or $J/\psi \psi(2S)$) are significantly smaller than the experimental widths of $X(6900)$ and $X(7200)$, which are around $100$ MeV~\cite{LHCb:2020bwg,CMS:2023owd,ATLAS:2023bft}. Therefore, these narrow P-wave resonances do not match well with the experimental states and remain to be discovered. They might be hidden within the broad structures of S-wave states in the $J/\psi J/\psi$ or $J/\psi \psi(2S)$ decay channels.}

    \clabel[Q2]{We find that the lowest resonant state in both S-wave and P-wave systems are located around $ 7 $ GeV, approximately $1$ GeV higher than the constituent mass $4m_c\approx6$ GeV.  This is reasonable since the resonant states lie above the di-charmonia thresholds, and the masses of the excited charmonia exceed $2m_c\approx 3$ GeV due to the confinement potential.} The broad structure ranging from $ 6.2 $ to $ 6.8 $ GeV in the experiments remains elusive and unseen in the quark model calculations. One possible explanation could be that the structure arises from extremely broad states. In this work and our previous work on S-wave systems~\cite{Wu:2024euj}, we only focus on states with widths $ \Gamma<200 $ MeV. The wave functions of states with larger widths are more divergent, and larger complex scaling angles are needed for numerically stable results, which are difficult to achieve with the current level of precision. Therefore, if $ X(6400) $ and $ X(6600) $ are states with widths larger than $ 200 $ MeV, as suggested by the experimental results~\cite{ATLAS:2023bft,CMS:2023owd}, they cannot be obtained numerically in the current calculations. Other possibilities include modifications of confinement mechanism in the quark model~\cite{Wang:2023jqs}, or contributions from dynamical effects rather than tetraquark resonant poles~\cite{Wang:2020wrp}.

	\section{Summary}\label{sec:summary}
	We calculate the energy spectrum of P-wave fully charmed tetraquark systems within the quark potential model. The two-body potential consists of the central term, spin-orbit term and tensor term, which arise from the one-gluon-exchange and linear confinement interactions. We diagonalize the complete Hamiltonian instead of treating the spin-orbit and tensor terms as perturbations. We solve the four-body Schr\"odinger equation using the Gaussian expansion method, taking both dimeson and diquark-antidiquark spatial configurations into account. We employ the complex scaling method to distinguish tetraquark resonant states from meson-meson scattering states.
 	
 	We obtain fully charmed P-wave tetraquark resonant states in the mass region of $(7.0,7.2)$ GeV. From the root-mean-square radii, we find that they all have compact tetraquark configurations, \clabel[Q3.2]{similar to the S-wave fully heavy tetraquark states in our previous studies~\cite{Wu:2024euj,Wu:2024hrv}.}
    The color mixing effect between $\chi_{\bar{3}_c\otimes3_c}$ and $\chi_{6_c\otimes \bar6_c}$ configurations plays a vital role in the resonant states, while the mixing between different spin configurations $\chi^S$ is very weak. Tetraquark resonant states with exotic quantum numbers $J^{PC}=0^{--}$ and $1^{-+}$ lie around $7.1$ GeV. They can be searched for in future experiments.
 	
 	\clabel[Q3.3]{In our previous work~\cite{Wu:2024euj}, we investigated the S-wave fully charmed tetraquark system using the same quark potential model. The theoretical mass spectrum of the S-wave and P-wave fully charmed tetraquark states are compared with the experimental results in Fig.~\ref{fig:spectrum4c}. In S-wave systems, 
    good candidates for experimentally observed $X(6900)$ and $X(7200)$ are obtained in the $J^{PC}=0^{++}$ and $2^{++}$ systems. However, 
    the P-wave resonances that can decay into the experimentally observed modes ($J/\Psi J/\Psi$ or $J/\Psi\Psi(2S)$) are rather narrow and do not match well with experimentally observed states. These states await experimental discoveries in the near future.} The broad structure ranging from $ 6.2 $ to $ 6.8 $ GeV in the experiments remains absent in both S-wave and P-wave systems in the quark model calculations. If the structure arises from resonant states with widths larger than $200$ MeV, they cannot be obtained numerically in the current calculations. Further theoretical studies and experimental measurements are needed to understand $X(6400)$ and $X(6600)$ better.
 	
	\section*{ACKNOWLEDGMENTS}
	
	We thank Dr. Lu Meng, Yan-Ke Chen, Dr. Yao Ma, Dr. Zi-Yang Lin, Dr. Jun-Zhang Wang and Liang-Zhen Wen for the helpful discussions. This project was supported by the National
	Natural Science Foundation of China (No. 12475137). The computational resources were supported by High-performance Computing Platform of Peking University.
	
	\newpage
	\appendix
	\section{Matrix element}
    \label{appsec:me}
	Direct evaluation of the spatial matrix elements of P-wave states involves complicated integrals of spherical harmonics of different configurations. To avoid this, we make use of the infinitesimally-shifted Gaussian basis functions~\cite{HIYAMA2003223}. The P-wave Gaussian basis can be expressed as
	\begin{equation}
		re^{-\nu r^2}Y_{1m}(\hat{r})=\boldsymbol{k}\cdot\boldsymbol{r}e^{-\nu r^2}=\lim_{\epsilon\rightarrow0}\frac{1}{4\epsilon\nu}[e^{-\nu(\boldsymbol{r}-\epsilon \boldsymbol{k})^2}-e^{-\nu(\boldsymbol{r}+\epsilon \boldsymbol{k})^2}],
	\end{equation}
	with
	\begin{equation}
		\boldsymbol{k}=\left\{
		\begin{array}{ll}
			\left(-\sqrt{\frac{3\pi}{8}},-\sqrt{\frac{3\pi}{8}}i,0\right),&m=1\\
			\left(0,0,\sqrt{\frac{3\pi}{4}}\right),& m=0\\
			\left(\sqrt{\frac{3\pi}{8}},-\sqrt{\frac{3\pi}{8}}i,0\right),&m=-1\\
		\end{array} \right. .
	\end{equation}
	
	To evaluate matrix elements of the central potential, we need to carry out the following type of integrals,
	\begin{widetext}
	\begin{equation}\label{appeq:cenint}
		\lim_{\epsilon_a,\epsilon_b\rightarrow0}\frac{1}{\epsilon_a\epsilon_b}\int d\mathcal{R}_ce^{-(\mathcal{R}_a^T-\mathcal{K}_a^T\epsilon_a)\nu_a(\mathcal{R}_a-\epsilon_a\mathcal{K}_a)}f(r_{c3})e^{-(\mathcal{R}_b^T-\mathcal{K}_b^T\epsilon_b)\nu_b(\mathcal{R}_b-\epsilon_b\mathcal{K}_b)},
	\end{equation}
	\end{widetext}
	with $\mathcal{R}_{\alpha}=\left(\boldsymbol{r}_{\alpha1},\boldsymbol{r}_{\alpha2},\boldsymbol{r}_{\alpha3}\right)$ being a set of Jacobian coordinates. By carrying out the transformation $\mathcal{R}_a=U_{ac}\mathcal{R}_c, \mathcal{R}_b=U_{bc}\mathcal{R}_c$, the integral can be expressed as
	\begin{equation}\label{appeq:Gausint1}
		\begin{aligned}
		&\int d\mathcal{R}_ce^{-(\mathcal{R}_c^T-\mathcal{D}^T)\mathcal{A}(\mathcal{R}_c-\mathcal{D})-\mathcal{C}}f(r_{c3})\\
		=&\int d\mathcal{R}_ce^{-(\mathcal{R}_c^T-\mathcal{D}^T)\mathcal{B}^T\mathcal{B}(\mathcal{R}_c-\mathcal{D})-\mathcal{C}}f(r_{c3})\\
		=&\operatorname{det}(\mathcal{B}^{-1})^3\int d\mathcal{\tilde{R}}_ce^{-(\mathcal{\tilde{R}}_c-\mathcal{\tilde{D}})^2-\mathcal{C}}f(\mathcal{B}^{-1}_{33}\tilde{r}_{c3}),\\
		\end{aligned}
	\end{equation}
	where $\mathcal{A}$ is a real symmetric matrix. In the second line, the Cholesky decomposition $\mathcal{A}=\mathcal{B}^T\mathcal{B}$ is used, with $\mathcal{B}$ being an upper triangular matrix. In the last line, $\mathcal{R}_c,\mathcal{D}$ are transformed as $\mathcal{\tilde{R}}_c=\mathcal{B}\mathcal{R}_c,\mathcal{\tilde{D}}=\mathcal{B}\mathcal{D}$. The Gaussian integral~\eqref{appeq:Gausint1} can be carried out analytically, which are kept up to terms linear in $\epsilon_a\epsilon_b$ to give the final results of Eq.~\eqref{appeq:cenint}.
	
	The spatial part of the tensor potential takes the form,
	\begin{equation}
		V_{\rm ts}\sim \frac{r_ir_j}{r^5},
	\end{equation}
	while for the spin-orbit potential, the orbital angular momentum operator is written as 
	\begin{equation}
		L_i=\epsilon_{ijk}r_jp_j=-i\epsilon_{ijk}r_j\frac{\partial}{\partial r_k}.
	\end{equation}
	The derivative acts on the Gaussian basis to give a constant or $r_k$ factor. Therefore, we need to carry out the integrals of the form,
	\begin{equation}\label{appeq:Gausint2}
		\begin{aligned}
			&\int d\mathcal{\tilde{R}}_ce^{-(\mathcal{\tilde{R}}_c-\mathcal{\tilde{D}})^2-\mathcal{C}}f(\tilde{r}_{c3})\tilde{r}_{c3,j},\\
			&\int d\mathcal{\tilde{R}}_ce^{-(\mathcal{\tilde{R}}_c-\mathcal{\tilde{D}})^2-\mathcal{C}}f(\tilde{r}_{c3})\tilde{r}_{c3,j}\tilde{r}_{c3,k},
		\end{aligned}
	\end{equation}
	where $\tilde{r}_{c3,i}$ denotes the $i$-th component of the coordinate $\boldsymbol{\tilde{r}}_{c3}$. The integrals in Eq.~\eqref{appeq:Gausint2} can be related to the integral in Eq.~\eqref{appeq:Gausint1} by taking derivatives with respect to $\mathcal{\tilde{D}}_{3,j}$ and $\mathcal{\tilde{D}}_{3,k}$,
	\begin{equation}
		\begin{aligned}
			&\int d^3 r f(r)e^{-(\boldsymbol{r}-\boldsymbol{d})^2} r_j=e^{-d^2}\frac{\partial}{2\partial d_j}\int d^3 r f(r)e^{-r^2+2\boldsymbol{d}\cdot\boldsymbol{r}},\\
			&\int d^3 r f(r)e^{-(\boldsymbol{r}-\boldsymbol{d})^2} r_j r_k=e^{-d^2}\frac{\partial}{2\partial d_j}\frac{\partial}{2\partial d_k}\int d^3 r f(r)e^{-r^2+2\boldsymbol{d}\cdot\boldsymbol{r}}.\\
		\end{aligned}
	\end{equation}
	Using the analytical solutions of Eq.~\eqref{appeq:Gausint1}, the integrals in Eq.~\eqref{appeq:Gausint2} can also be carried out analytically.

	\bibliography{p4cref}
	
	\onecolumngrid
	\clearpage
	\twocolumngrid

\end{document}